\documentclass[preprint,aps]{revtex4-1}
\usepackage{hyperref}
\usepackage{graphicx}
\usepackage{epstopdf}
\usepackage{subfig}
\usepackage{bm}
\usepackage{color}
\usepackage{soul}
\usepackage{slashed}
\usepackage{caption}
\usepackage{float}
\def\be{\begin{equation}}
\def\ee{\end{equation}}
\def\bea{\begin{eqnarray}}
\def\eea{\end{eqnarray}}
\def\ba{\begin{array}}
\def\ea{\end{array}}
\def\bc{\begin{center}}
\def\ec{\end{center}}
\begin{document}
\title{Transverse distortion of a relativistic composite system in impact parameter space}
\author{Narinder Kumar and Harleen Dahiya}
\address{Department of Physics\\
Dr. B. R. Ambedkar National Institute of Technology\\
         Jalandhar-144011, India}
\begin{abstract}
We investigate the Generalized Parton Distributions (GPDs) in impact parameter space using the explicit light front wave functions (LFWFs) for the two-particle Fock state of the electron in QED. The Fourier transform (FT) of the GPDs gives the distribution of quarks in the transverse plane for zero longitudinal momentum transfer ($\xi=0$). We study the relationship of  the spin flip GPD $E(x,0,-\vec{\Delta}_\perp^2)$ with the distortion of unpolarized quark distribution in the transverse plane when the target nucleon is transversely polarized and also determine the sign of distortion from the sign of anomalous magnetic moment. To verify the sign of distortion, we also compute it directly from the LFWFs by performing a FT in position space coordinate $\vec{f}_\perp$. The explicit relation between the deformation in the two spaces can also be obtained using the convolution integrals.
To show the relation of the model LFWFs  to a realistic model of nucleon physics, we have  designed a specific weight function of our model LFWFs and integrated it over the mass parameter. Also we have simulated the form factor of the nucleon in the  AdS/QCD holographic LFWFs model and studied the power-law behaviour at short distances.

\end{abstract}
\maketitle

\section{Introduction}
Deep virtual compton scattering (DVCS) \cite{dvcs,dvcs1,dvcs2,dvcs3} is the main process to probe the internal structure of hadrons.  Recently, the Generalized Parton Distributions (GPDs) \cite{gpds,gpds1,gpds2,gpds3,gpds4,miller1,miller2,miller3,miller4,miller5,pire} have attracted a considerable amount of interest towards this. GPDs  allow us to access partonic configurations not only with a given longitudinal momentum fraction but also at a specific (transverse) location inside the hadron. GPDs can be related to the angular momentum carried by quarks in the nucleon and the distribution of quarks can be described in the longitudinal direction as well as in the impact parameter space \cite{longit,imp2,imp1,imp3,imp4,imp0}.

When integrated over $x$ the GPDs reduce to the form factors which are the non-forward matrix element of the current operator and they describe how the forwards matrix element (charge) is distributed in position space. The GPDs are the off-forward matrix elements and it is well known that they reduce to Parton Distribution Functions (PDFs) in the forward limit. On the other hand, Fourier transform (FT) of GPDs w.r.t. transverse momentum transfer gives the distribution of partons in transverse position space \cite{imp2,imp1}. Therefore, their should be some connection between transverse position of partons and FT of GPDs w.r.t. transverse momentum transfer.


With the help of impact parameter dependent parton distribution function (ipdpdf) one can obtain the transverse position of partons in the transverse plane. However, it is not possible to measure the longitudinal position of partons. In order to measure the transverse position with the  longitudinal momentum simultaneously we can consider the polarized nucleon state in the transverse direction which leads to  distorted unpolarized ipdpdf in the tranverse plane \cite{ipd_buk1,ipd_buk2,ipd11,ipd12}.
Distortion obtained in the transverse plane also leads to single spin asymmetries (SSA) and it has been shown that such asymmetries can be explained by final state interactions (FSI) \cite{ssa,ssa1,fsi}. This mechanism gives us a good interpretation of SSAs which arises from the asymmetry (left-right) of quarks distribution in impact parameter space.

To study the GPDs, we use light front wave functions (LFWFs) which give a very simple representation of GPDs.
Impact parameter dependent parton distribution functions have been investigated by using the explicit LFWFs for the two-particle Fock state of the electron in QED \cite{model1,kumar1,kumar2,kumar3}.
In the present study we use the model consisting of spin-$\frac{1}{2}$ system as a composite of spin-$\frac{1}{2}$ fermion and spin-1 vector boson.  We have generalized the
framework of QED by assigning a mass $M$ to external electrons in the Compton scattering process, but a different mass $m$ to the internal electron
line and a mass $\lambda$ to the internal photon line. The idea behind this is to model the structure of a composite fermion state with a mass $M$ by a fermion and a vector constituent with respective masses $m$ and $\lambda$ \cite{model,model1}. In our case we take $\xi=0$ \cite{ipd2,ipd3} which represents the momentum transfer exclusively in transverse direction leading to the study of ipdpdfs in transverse impact parameter space.
In order to show the relation of the LFWFs in the two-particle Fock state of the electron in QED to a realistic model of nucleon physics, we have  designed a specific weight function of our model LFWFs and integrated it over the mass parameter. The Dirac and Pauli form factors have been simulated to obtain the correct perturbative QCD fall-off of the form factors at large $q^2$.
Also we have simulated the form factor of the nucleon in the  AdS/QCD holographic LFWFs model \cite{ads_qcd,ads_elec_gravit,ads_gravit} and studied the power-law behaviour of wavefunction at short distances.

For the case of spin flip GPD $E(x,0,-\vec{\Delta}_\perp^2)$, the parton distribution is distorted in the transverse plane when the target has a transverse polarization and when integrated over $x$, $E(x,0,-\vec{\Delta}_\perp^2)$ yields the Pauli form factor $F_2(t)$. The study of the Fourier transformed GPD $\mathcal{E}(x,b_\perp)$ is important for a transversely polarized target since it measures the distortion of the parton distribution in the transverse plane. Integrating ipdpdf $\mathcal{E}(x,b_\perp)$ over $b_\perp$ and $x$ gives us the magnetic moment. The sign of distortion can be concluded from the sign of the magnetic moment of the nucleon.
We extend the calculations to determine this sign of distortion from the  unintegrated momentum space distribution obtained directly from the LFWFs which can be obtained after performing a FT to relative position space coordinate $\vec{f}_\perp$. This is the another direct way to determine the sign of distortion from the LFWFs. The explicit relation between the deformation calculated from GPDs in the impact parameter space and the deformation calculated directly from the LFWFs can also be obtained.
\section{Generalized Parton Distributions (GPDs) }
The GPDs $H,E$ are defined through matrix elements of the bilinear vector currents on the light cone \cite{gpds2,gpds5,imp1}:
\bea
&& \int \frac{dy^-}{8 \pi} e^{i x P^+ y^-/2} \langle P'|\bar{\psi}(0) \gamma^+ \psi(y)| P\rangle|_{y^+=0,y_\perp=0} \nonumber\\
&=&\frac{1}{2 \bar{P^+}} \bar{U}(P')[H(x,\xi,t) \gamma^+ +E(x,\xi,t) \frac{i}{2M} \sigma^{+ \alpha}(\Delta_\alpha)] U(P).
\label{e1}
\eea
Here, $\bar{P}=\frac{1}{2}(P'+P)$ is the average momentum of the initial and final hadron and $\xi$ is the skewness parameter. Since we are considering the case where momentum transfer is purely transverse, we take the skewness parameter $\xi=0$ and in that case $t= - \vec{\Delta}_\perp^2$ is the invariant momentum transfer.
The off-forward matrix elements can be expressed as overlaps of the light front wave functions (LFWFs) for the two-particle Fock state of the electron in QED. We consider here a spin-$\frac{1}{2}$ system as a composite of spin-$\frac{1}{2}$ fermion and spin-1 vector boson. The details of the model have been presented in Ref. \cite{model1}, however, for the sake of completeness we present here the essential two-particle wave functions for spin up and spin down electron  expressed as
\begin{eqnarray}
&&\psi_{+\frac{1}{2}+1}^{\uparrow}(x,\vec{k}_\perp)=-\sqrt{2}\frac{-k^1+i k^2}{x(1-x)}\varphi,\nonumber\\ && \psi_{+\frac{1}{2}-1}^{\uparrow}(x,\vec{k}_\perp)=-\sqrt{2}\frac{k^1+ ik^2}{(1-x)}\varphi,\nonumber\\ &&
\psi_{-\frac{1}{2}+1}^{\uparrow}(x,\vec{k}_\perp)=-\sqrt{2}\left(M-\frac{m}{x}\right)\varphi,\nonumber\\ &&\psi_{-\frac{1}{2}-1}^{\uparrow}(x,\vec{k}_\perp)=0 \,, \label{spinup}
\end{eqnarray}
and
\begin{eqnarray}
&&\psi_{+\frac{1}{2}+1}^{\downarrow}(x,\vec{k}_\perp)=0,\nonumber\\&&
\psi_{+\frac{1}{2}-1}^{\downarrow}(x,\vec{k}_\perp)=-\sqrt{2}\left(M-\frac{m}{x}\right)\varphi,\nonumber\\ &&\psi_{-\frac{1}{2}+1}^{\downarrow}(x,\vec{k}_\perp)=-\sqrt{2}\frac{-k^1+i k^2}{(1-x)}\varphi,\nonumber\\
&&\psi_{-\frac{1}{2}-1}^{\downarrow}(x,\vec{k}_\perp)=-\sqrt{2}\frac{k^1+i k^2}{x(1-x)}\varphi \,,
\label{spindown}
\end{eqnarray}
where
\begin{eqnarray}
\varphi(x, \vec{k}_{\perp}) =\frac{e}{\sqrt {1-x}} \frac{1}{M^2-\frac{\vec{k}^2_{\perp}+m^2}{x}-\frac{\vec{k}_{\perp}^{2}+\lambda^2}{1-x}}\,.
\end{eqnarray}
The framework of QED has been generalized by assigning a mass $M$ to external electrons in the Compton scattering process, but a different mass $m$ to the internal electron line and a mass $\lambda$ to the internal photon line.
Using the above wavefunctions, the helicity non-flip and flip GPDs can be expressed as
\bea
H(x,0,-\vec{\Delta}_\perp^2)&=& \int{\frac{d^2\vec{k}_\perp}{16\pi^3}} \bigg[ \psi_{+\frac{1}{2}+1}^{\uparrow *}(x,\vec{k}'_\perp) \psi_{+\frac{1}{2}+1}^{\uparrow}(x,\vec{k}_\perp)+\psi_{+\frac{1}{2}-1}^{\uparrow *}(x,\vec{k}'_\perp) \psi_{+\frac{1}{2}-1}^{\uparrow}(x,\vec{k}_\perp)+\nonumber\\
&& \psi_{-\frac{1}{2}+1}^{\uparrow *}(x,\vec{k}'_\perp) \psi_{-\frac{1}{2}+1}^{\uparrow}(x,\vec{k}_\perp) \bigg] \,,
\label{h2}
\eea
\be
\frac{\Delta^1- i \Delta^2}{2 M} E(x,0,-\vec{\Delta}_\perp^2)=\int{\frac{d^2\vec{k}_\perp}{16\pi^3}} \Big[\psi_{+\frac{1}{2}+1}^{\uparrow *}(x,\vec{k}'_\perp) \psi_{+\frac{1}{2}+1}^{\downarrow}(x,\vec{k}_\perp)+\psi_{+\frac{1}{2}-1}^{\uparrow *}(x,\vec{k}'_\perp) \psi_{+\frac{1}{2}-1}^{\downarrow}(x,\vec{k}_\perp) \Big].
\label{e2}
\ee
Using eqs. (\ref{spinup}) and (\ref{spindown}) as well as the relation $\vec{k}'_\perp=\vec{k}_\perp-(1-x)\vec{\Delta}_\perp$, we get
\be
E(x,0,-\vec{\Delta}_\perp^2)=- 2 M \left(M-\frac{m}{x}\right)x^2(1-x) I_1 ,
\label{e3}
\ee
where
\be
I_1= \pi \int_{0}^{1} \frac{d \alpha}{D} ,
\label{e4}
\ee and
\be
D= \alpha (1-\alpha) (1-x)^2 \Delta^2_{\perp}-M^2 x (1-x)+ m^2 (1-x)+\lambda^2 x .
\label{e5}
\ee
Since the FT diagonalizes the convolution integral, we switch to transverse position space representation of the LFWF by taking FT in $\vec{\Delta}_\perp$ as
\bea
\mathcal{H}(x,\vec{b}_\perp)&=&\frac{1}{(2 \pi)^2}\int d^2\vec{\Delta}_\perp e^{-i \vec{b}_\perp \cdot \vec{\Delta}_\perp} H(x,0,-\vec{\Delta}_\perp^2)=\frac{1}{2 \pi}\int \Delta \ d {\Delta} J_0(\Delta b) H(x,0,-\vec{\Delta}_\perp^2),\nonumber\\
\mathcal{E}(x,\vec{b}_\perp)&=&\frac{1}{(2 \pi)^2}\int d^2\vec{\Delta}_\perp e^{-i \vec{b}_\perp \cdot \vec{\Delta}_\perp} E(x,0,-\vec{\Delta}_\perp^2)=\frac{1}{2 \pi}\int \Delta \ d {\Delta} J_0(\Delta b) E(x,0,-\vec{\Delta}_\perp^2),
\label{fourierHE}
\eea
where $J_0(\Delta b)$ is the Bessel function and $\vec{b}_\perp$ is the impact parameter conjugate to $\vec{\Delta}_\perp$ representing the transverse distance between the active quark and the center of mass momentum.
\begin{figure}
\minipage{0.42\textwidth}
    \includegraphics[width=5.8cm,angle=270]{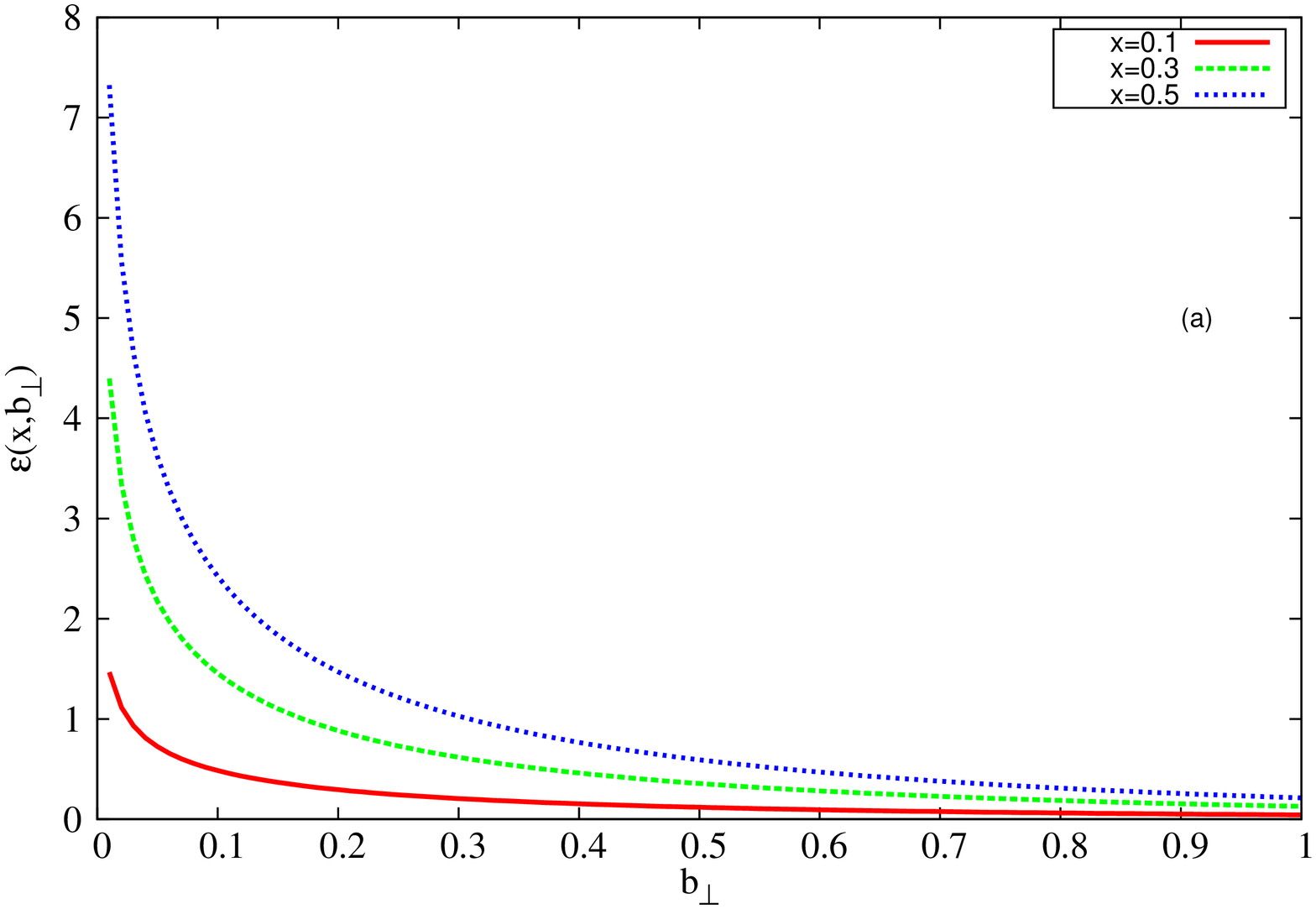}
  \endminipage\hfill
  \minipage{0.42\textwidth}
  \includegraphics[width=5.8cm,angle=270]{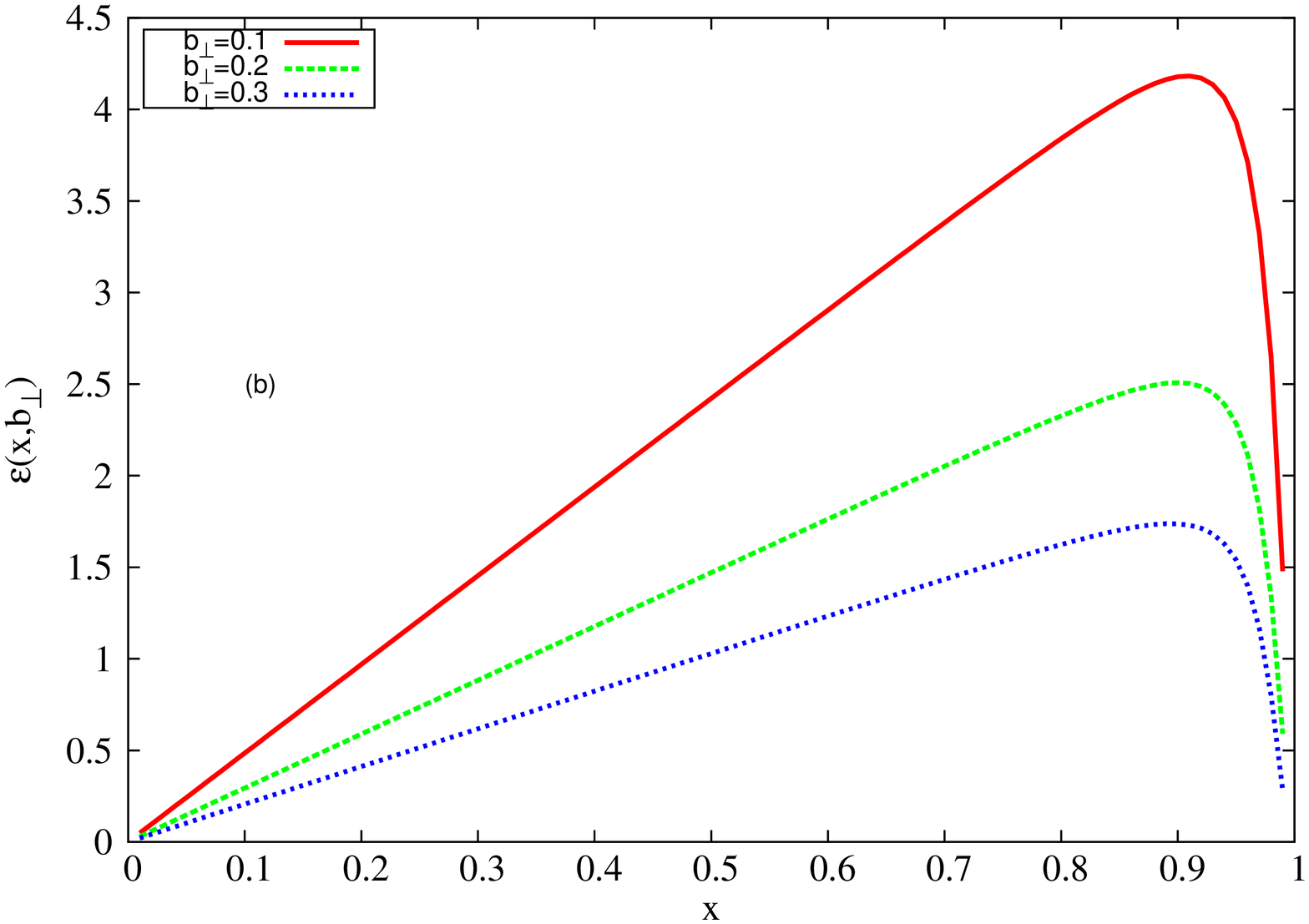}
  \endminipage\hfill
 \caption{Plots of $\mathcal{E}(x,b_\perp)$ as a function of $ b_\perp $ and $x$ for three different values of $x$ and $b_\perp$ respectively.}
  \label{impactE}
\end{figure}\\
\begin{figure}
\minipage{0.42\textwidth}
   \includegraphics[width=6cm]{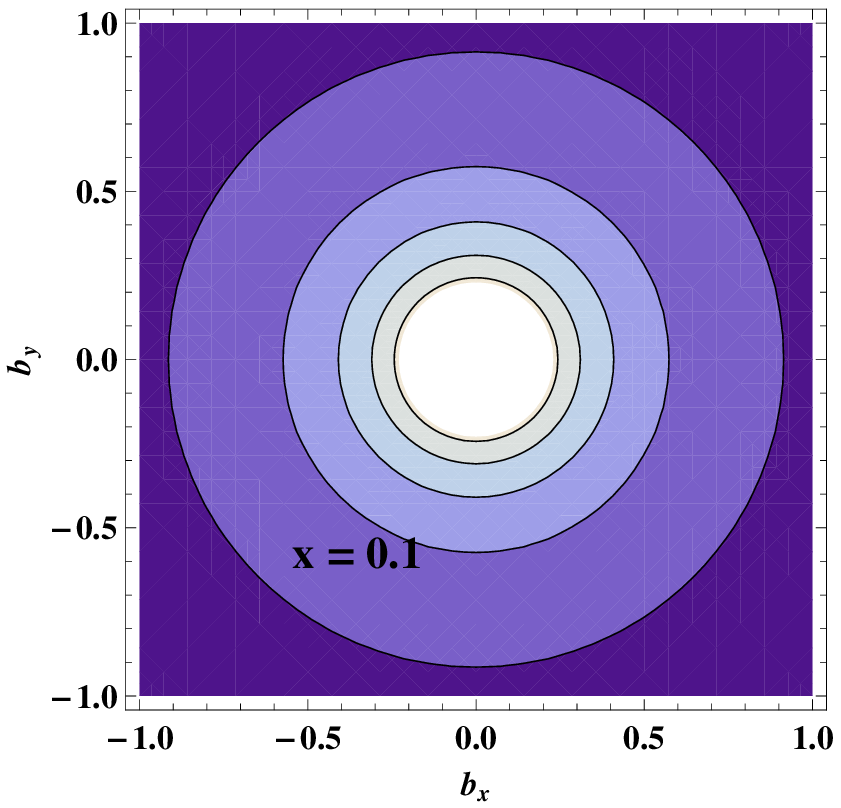}
  \endminipage\hfill
  \minipage{0.42\textwidth}
  \includegraphics[width=6cm]{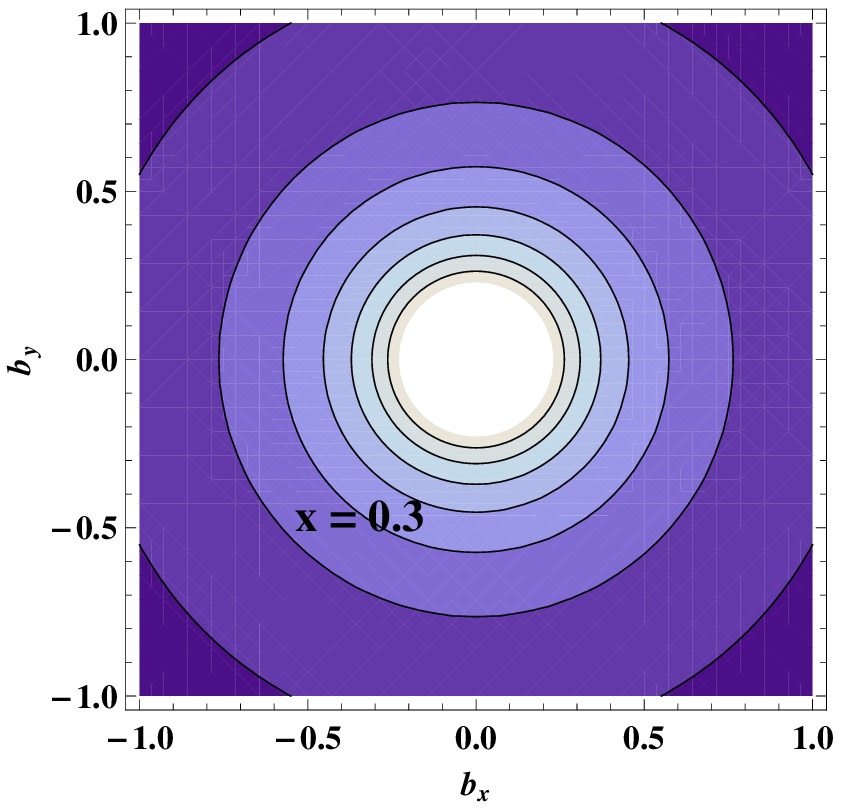}
  \endminipage\hfill
  \minipage{0.42\textwidth}
  \includegraphics[width=6cm]{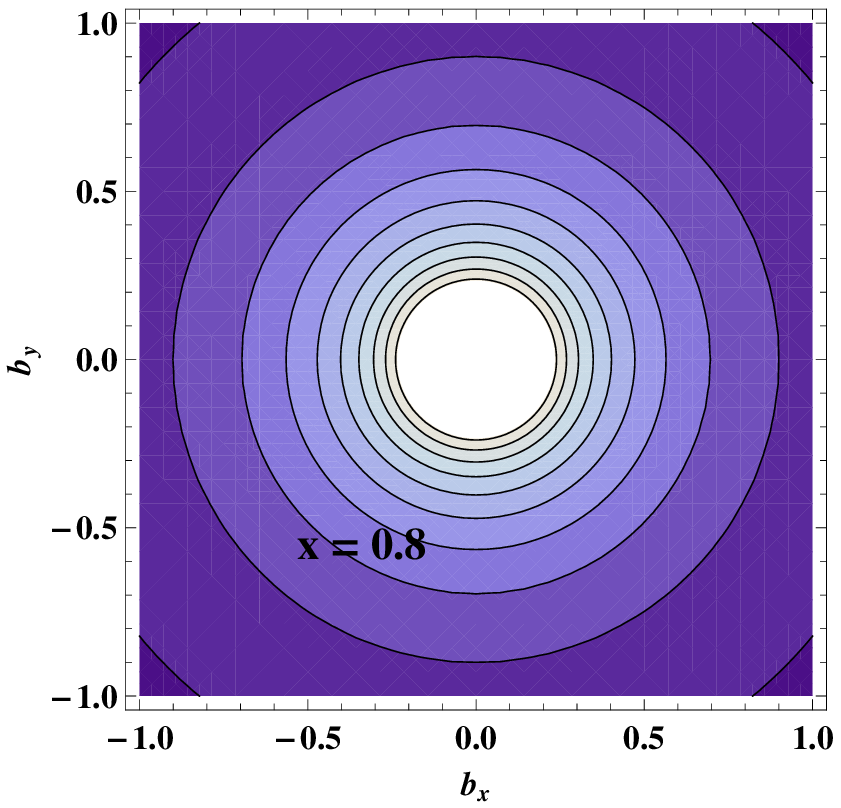}
  \endminipage\hfill
  \caption{Plots of ipdpdf $\mathcal{E}(x,b_\perp)$ for three different values of $x$.}
  \label{contour}
\end{figure}
In fig. \ref{impactE}(a), we present the impact parameter dependent parton distribution function $\mathcal{E}(x,b_\perp)$ as a function of ${b_\perp}$ for different fixed values of $x$. We have taken $M=0.5$MeV, $m=0.5$MeV and $\lambda=0.02$MeV for our numerical calculations. We can see that $\mathcal{E}(x,\vec{b}_\perp)$ decreases as $\vec{b}_\perp$ increases. Since $\xi=0$ in the present study, it clearly implies that there is no finite momentum transfer in the longitudinal direction. This in turn suggests that the initial and final transverse positions of the proton remain the same and the probability interpretation is  now possible. It is important to note here that $\mathcal{E}(x,\vec{b}_\perp)$ for a free Dirac particle is a delta function and the smearing observed in the $|{b_\perp}|$ space is due to the spin correlation in the two particle LFWFs. It is clear from the plots that the partons are distributed mostly near $b_\perp=0$ which is the center of momentum. As we move away from the center of momentum towards larger values of $b_\perp$, the density of partons decreases.

Further, the magnitude of $\mathcal{E}(x,\vec{b}_\perp)$ increases with the increasing value of the momentum fraction $x$. In fig. \ref{impactE}(b), we have plotted the $\mathcal{E}(x,b_\perp)$ as a function of $x$ with three different values of $b_\perp$. It can be clearly seen that it increases as the value of $x$ increases and tends to zero at $x \rightarrow 1$. In fig. \ref{contour} we present quark distribution in the transverse plane for $x$= 0.1, 0.3 and 0.8. It describes the quark distribution for the unpolarized nucleon and it is clear from the plot that for $x$=0.1 the distribution is spread over the whole region but as the value of $x$ increases it gets denser near the center.

\begin{figure}
\minipage{0.42\textwidth}
    \includegraphics[width=8cm,angle=0]{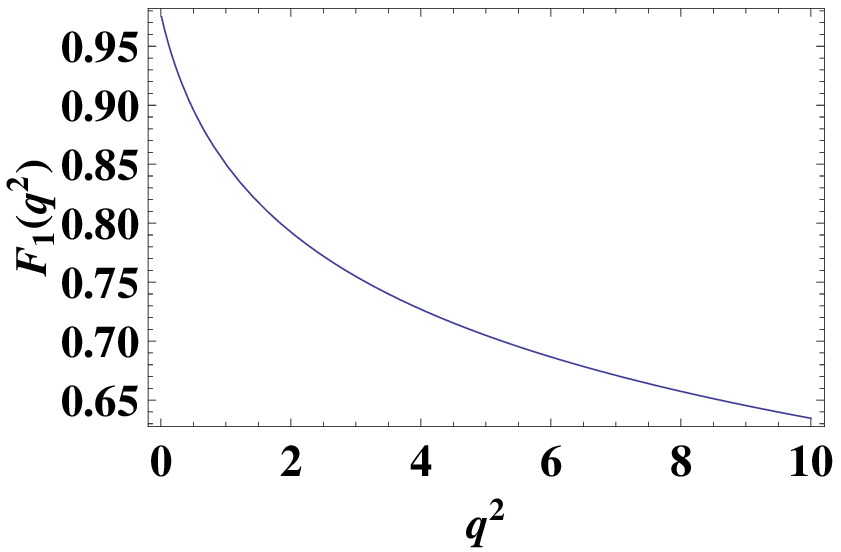}
  \endminipage\hfill
  \minipage{0.42\textwidth}
  \includegraphics[width=8cm,angle=0]{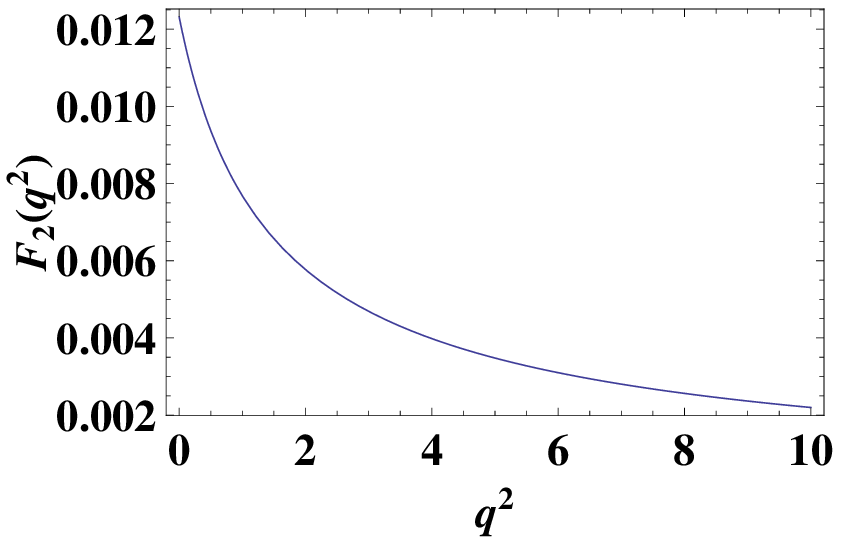}
  \endminipage\hfill
 \caption{ Plots of simulated Dirac and Pauli form factors as a function of $q^2$ after integrating over $M^2$.}
  \label{simulated}
\end{figure}

In order to have a deep insight of this model in context of the well know nucleon properties, we
design the model of LFWFs integrated over the mass parameter
\be
\int dM^2 \rho(M^2) M^2,
\label{simulate}
\ee
where $\rho(M^2)$ is the weight function. We have chosen $\rho(M^2)= e^{- \frac{M^2}{\Lambda^2_{QCD}(1-x)^2}}$, which is not only consistent with the $x \rightarrow 1$ and $x \rightarrow 0$ constraints but also its integration gives the correct perturbative QCD fall-off of the Dirac and Pauli form factors at large $q^2$. We would like to emphasize here that even though earlier studies have already produced this behavior without any weight function \cite{ma-ivan}, the ad hoc weighting function is introduced so that a relation to a realistic model of nucleon physics can be shown. The form factors obtained from LFWFs have been simulated using eq. (\ref{simulate}).  We have  taken an arbitrary parameter $y=x M^2$ to solve the integration given in eq. (\ref{simulate}) and the results for the Dirac and Pauli form factors have been obtained following Ref. \cite{model1}. In fig. \ref{simulated}, we have plotted the simulated Dirac and Pauli form factors as a function of $q^2$ and it is clear from the plots that, as expected, both form factors fall off at large $q^2$.

\begin{figure}
  \minipage{0.42\textwidth}
  \includegraphics[width=8cm,angle=0]{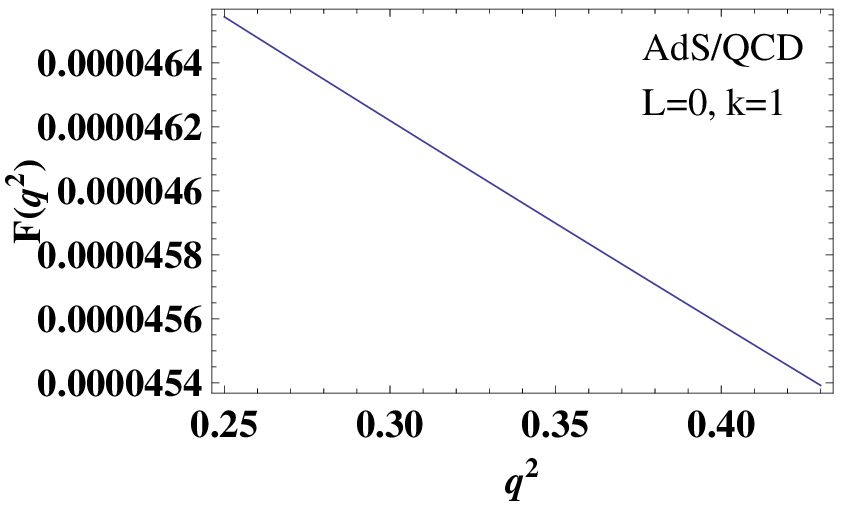}
  \endminipage\hfill
  \minipage{0.42\textwidth}
  \includegraphics[width=8cm,angle=0]{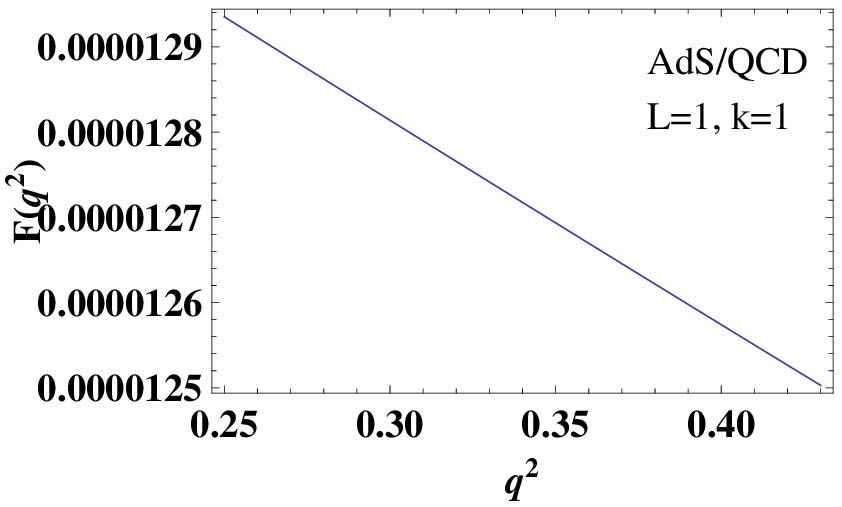}
  \endminipage\hfill
     \caption{Nucleon form factor as a function of $q^2$ for the ground ($L=0, k=1$) and the first orbital excited state ($L=1, k=1$) obtained from AdS/QCD model of LFWFs.}
  \label{ads}
\end{figure}

In addition to this,  we have simulated the form factor of the nucleon in the  AdS/QCD holographic LFWFs model \cite{ads_qcd,ads_elec_gravit,ads_gravit} where the LFWFs encode all the properties of hadron like bound state quark and gluon properties. The holographic model is quite successful in explaining the hadron spectrum and can act as template for composite systems describing the partonic structure.  Following Ref. \cite{ads_qcd}, we define a string amplitude $\Phi(z)$ on the fifth dimension in AdS$_5$ space which easily maps to the LFWFs of the hadrons and allow us to calculate the structure functions, form factors, DVCS constants etc.. Further, with $z\rightarrow 0$, the scale dependence determines the power-law behaviour of wavefunction at short distances and predicted behaviour matches with the available perturbative QCD results \cite{ads_pqcd}.  A correspondence exists between the fifth dimensional holographic variable $z$ and a impact separation variable $\zeta$.  The form factor in  AdS is given by the overlap of normalizable modes dual to the incoming and outgoing hadrons and  is given as
\be
F(q^2)=2 \pi \int_{0}^{1} \frac{dx}{x(1-x)} \int \zeta d\zeta J_0(\zeta \ q \ \sqrt{\frac{x}{1-x}}) |\Psi(x,\zeta)|^2,
\ee
where the normalized light front wavefunction for two particle state follows from \cite{ads_qcd}
\be
\Psi_{L,k}(x,\zeta)=B_{L,k} \sqrt{x (1-x)}J_L(\zeta \beta_{L,k} \Lambda_{QCD}) \theta(z\le \Lambda_{QCD}^{-1}),
\ee
where
\be B_{L,k}=\Lambda_{QCD} [(-1)^L \pi J_{1+L}(\beta_{L,k} \ \Lambda_{QCD}) J_{1-L}(\beta_{L,k} \ \Lambda_{QCD})]^{-\frac{1}{2}},
 \ee
and $\beta_{L,k}$ is the $k$th zero of the Bessel function $J_L$.
We have obtained the nucleon form factor from normalized LFWFs $\Psi_{L,k}$ as a function of $q^2$. In this case also, we have simulated the nucleon form factor by taking the integration over the parameter $M^2$ as described in eq. (\ref{simulate}). The form factor for the ground ($L=0, k=1$) and the first orbital excited state ($L=1, k=1$) are presented in fig. \ref{ads}. It is clear from the plots that the magnitude of the form factor falls-off at large value of $q^2$ and the light cone composite model used in the present work matches the power-law fall-off of form factors in perturbative QCD.

\section{Transverse distortion of the wave function}
To understand the physical significance of ipdpdf $\mathcal{E}(x,\vec{b}_\perp)$, we consider a state polarized in the $+\hat{y}$ direction with it's transverse center of momentum at the origin
\bea
|P^+,\vec{R}_\perp=\vec{0}_\perp,+\hat{y} \rangle=\frac{1}{\sqrt{2}}(|P^+,\vec{R}_\perp=\vec{0}_\perp,\uparrow \rangle +i |P^+,\vec{R}_\perp=\vec{0}_\perp,\downarrow \rangle,
\label{e6}
\eea
where
\bea
|P^+,\vec{R}_\perp=\vec{0}_\perp,\lambda \rangle= \mathcal{N} \int d^2\vec{P}_\perp |P^+,\vec{P}_\perp,\lambda \rangle.
\label{e7}
\eea
$\mathcal{N}$ is the normalization factor and it is chosen such that we get the parton distributions when the impact parameter dependent distributions are integrated over $d^2\vec{b}_\perp$.
The transverse distance from center of momentum can defined using the light cone  momentum density component of the energy momentum tensor and can be expressed as
\bea
\vec{R}_\perp\equiv \frac{1}{P^+} \int d^2\vec{x}_\perp \int dx^- T^{++} \vec{x}_\perp = \sum_{i=q,g} x_i \vec{r}_{\perp,i}.
\eea
Here, $x_i$ are the light cone momentum fractions carried by each parton and the sum in the parton representation of $\vec{R}_\perp$ extends over the transverse positions $\vec{r}_{\perp,i}$ of all quarks and gluons in the target. Using the operator
\bea
\hat{O}_q(x,\vec{b}_\perp)=\int \frac{dy^-}{4 \pi}\bar{\psi} \left( -\frac{y^-}{2} \vec{b}_\perp \right) \gamma^+ \psi \left( \frac{y^-}{2} \vec{b}_\perp \right) \times e^{ixP^+y^-},
\eea and the light front gauge $A^+=0$ for a state polarized in $+\hat{y}$ direction, we get the unpolarized quark distribution in impact parameter space \cite{ipd11,ipd12} expressed as
\bea
q_{\hat{y}}(x,\vec{b}_\perp)&=& \langle P^+, \vec{R}_\perp=\vec{0}_\perp, + \hat{y} | \hat{O}_q(x,\vec{b}_\perp)| P^+, \vec{R}_\perp=\vec{0}_\perp, +\hat{y} \rangle \nonumber\\
&=&\int \frac{d^2 \vec{\Delta}_\perp}{(2\pi)^2} e^{-i \vec{\Delta}_\perp \cdot \vec{b}_\perp} [H(x,0,-\vec{\Delta}_\perp^2)]+i \frac{\Delta^x}{2 M} E(x,0,-\vec{\Delta}_\perp^2)].
\label{e8}
\eea
Using eq. (\ref{fourierHE}), we get the unpolarized quark distribution in terms of the Fourier transformed GPDs as follows
\bea
q_{\hat{y}}(x,\vec{b}_\perp)&=& \mathcal{H}(x,\vec{b}_\perp)+\frac{1}{2 M} \frac{\partial}{\partial b^x} \mathcal{E}(x,\vec{b}_\perp).
\label{e9}
\eea
It is clear from the above expression that the parton distribution of quarks in the transverse plane is distorted for the target having transverse polarization when the $b^x$ derivative of $\mathcal{E}(x,b_\perp)$ is added provided the spin flip GPD $E(x,0,t)$ is non zero. On the one hand, integrating the spin-flip GPD $E(x,0,t)$  over $x$ gives the Pauli form factor $F_2(t)$ whereas on the other hand, integrating $\mathcal{E}(x,\vec{b}_\perp)$ over both $x$ and $\vec{b}_\perp$ gives the quark contribution to anomalous magnetic moment as follows
\bea
\int dx \int d^2 \vec{b}_\perp \mathcal{E}(x,\vec{b}_\perp)=\kappa.
\label{e10}
\eea
The sign of anomalous magnetic moment is important since it determines the sign of distortion of quark distribution in impact parameter space. It is well known that a Fourier transformed  function may have a maxima (minima) at the origin. From eq. (\ref{e9}) it can be clearly seen that when $\kappa$ is taken to be positive, the $b^x$ derivative of a smooth positive function $\mathcal{E}(x,b_\perp)$, with a maxima at the origin, is positive for negative $b^x$ and negative for positive $b^x$. The situation reverses for the negative values of $\kappa$. As a result, when $\kappa>0$, the nucleon which is polarized in the $\hat{y}$ direction, the distortion is towards negative $\hat{x}$ for positive $b^x$ and towards positive $\hat{x}$ for negative $b^x$. Similarly, when $\kappa<0$, the nucleon which is polarized in the $\hat{y}$ direction, the distortion is towards positive $\hat{x}$ for positive $b^x$ and towards negative $\hat{x}$ for negative $b^x$. We can verify the above assumptions in the present study.

\begin{figure}
\minipage{0.42\textwidth}
    \includegraphics[width=5.8cm,angle=270]{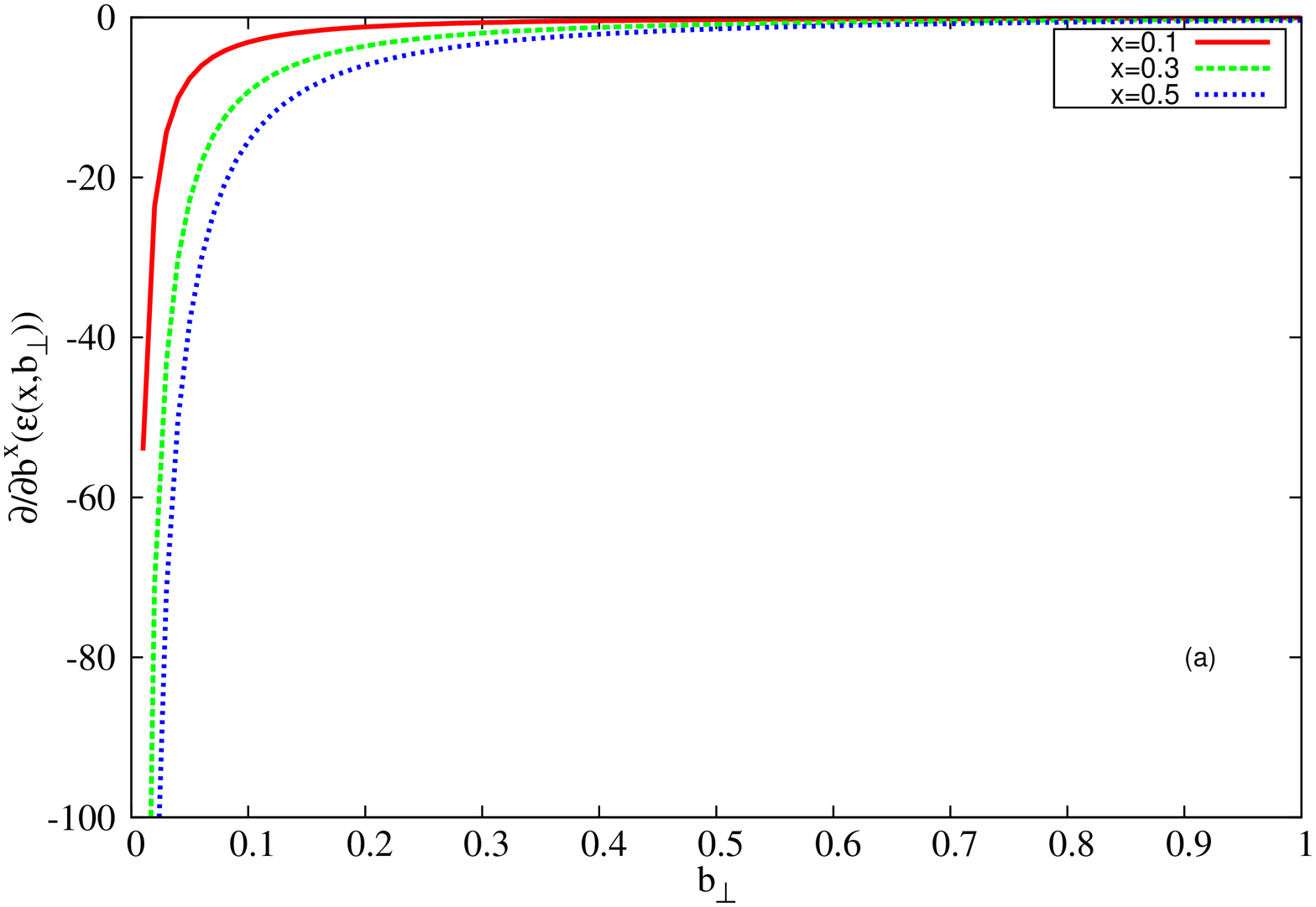}
  \endminipage\hfill
  \minipage{0.42\textwidth}
  \includegraphics[width=5.8cm,angle=270]{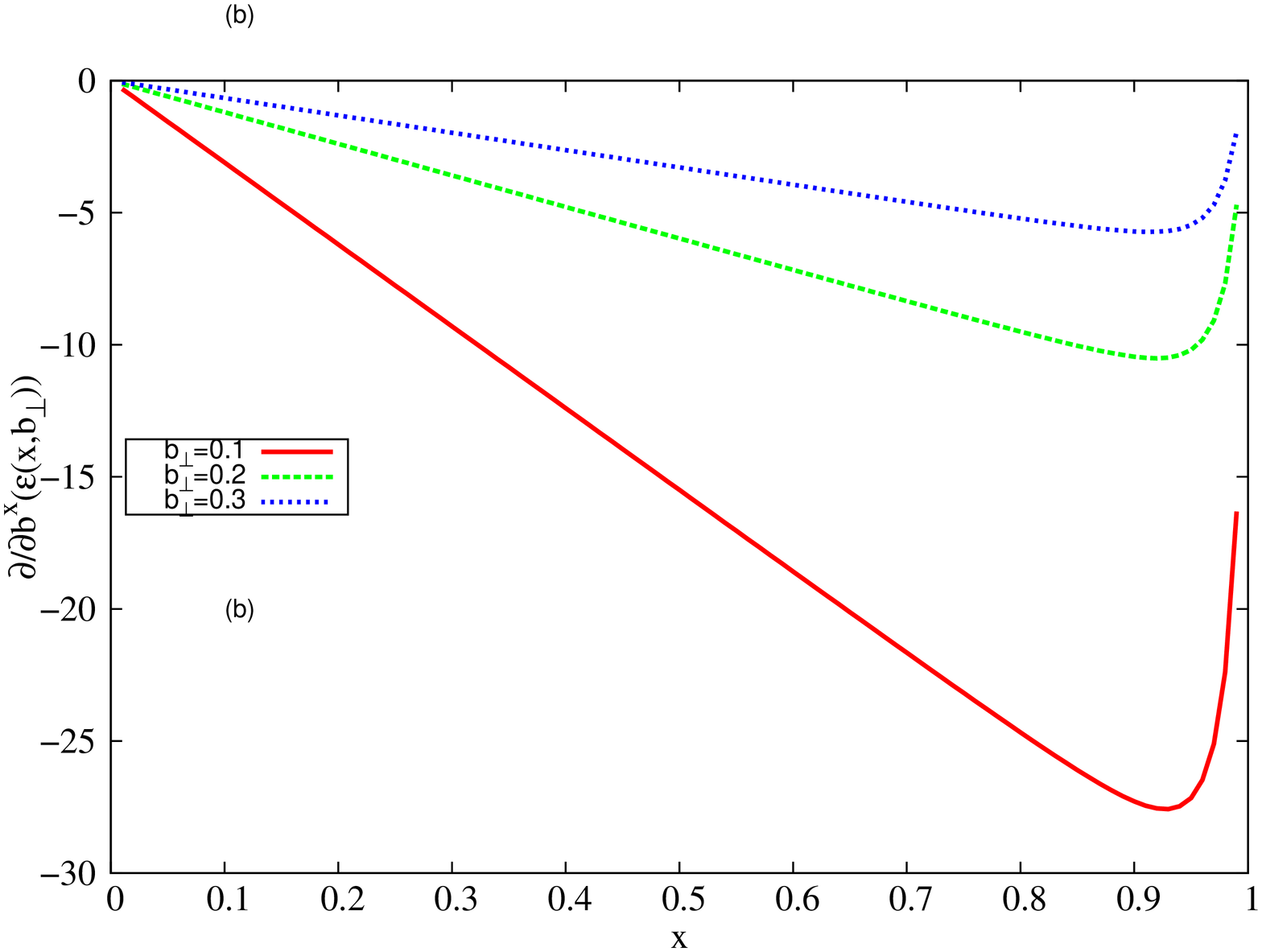}
  \endminipage\hfill
 \caption{Plots of $\frac{\partial}{\partial b^x} \mathcal{E}(x,b_\perp)$ as a function of $ b_\perp $ and $x$ for different values of $x$ and $b_\perp$ respectively.} \label{newx}
\end{figure}

The distortion in impact parameter space for a polarized nucleon in the present study is given as
\bea
\frac{\partial}{\partial b^x} \mathcal{E}(x,\vec{b}_\perp)&=& - \frac{1}{2 \pi} \int \Delta^2 J_1(\Delta b) E(x,0,-\vec{\Delta}_\perp^2) d\Delta .
\label{e11}
\eea
To have a deeper understanding we have plotted, in fig. \ref{newx}(a), the  distortion in impact parameter space $\frac{\partial}{\partial b^x} \mathcal{E}(x,b_\perp)$ for a nucleon polarized in the $\hat{y}$ direction  as a function of $b_\perp$. We have taken three different values of $x$ and it is clear from plot that magnitude of distortion increases as the value of $x$ increases. The distortion obtained is in negative direction because the anomalous magnetic moment is positive. This is in agreement with the results of Ref. \cite{ipd3} where a model of spin-$\frac{1}{2}$ system namely an electron dressed with a photon in QED has been used to study the distortion in impact parameter space. Again we have taken $M=0.5$MeV, $m=0.5$MeV and $\lambda=0.02$MeV for our numerical calculations. In fig. \ref{newx}(b), we present the $\frac{\partial}{\partial b^x} \mathcal{E}(x,b_\perp)$ vs $x$ for three different values of $b_\perp$. It represents the distortion of ipdpdf in the transverse plane for a transversely polarized target. It is clear from the plot that distortion increases as the value of $x$ increases but at $x \rightarrow 1$ it decreases. Further, the magnitude of distortion decreases as the value of $b_\perp$ increases. The study of transverse distortion is significant in the context of developing an intuitive explanation for transverse SSAs \cite{ipd11,ipd12}.

In addition to the sign of distortion in impact parameter space obtained above, there is infact an alternate way to determine this sign from the  unintegrated momentum space distribution obtained directly from the LFWFs. This can be achieved by performing a FT in position space coordinate $\vec{f}_\perp$. One can then explicitly show the relation between the deformation obtained from GPDs in the impact parameter space and as calculated directly from the LFWFs in the calculations presented below using the convolution integrals. These relations will also provide insight into the phenomena of shifting from the impact parameter space to the transverse position space representation. To this end, we start by taking the wavefunctions for a nucleon polarized in $+\hat{y}$ direction as follows
\bea
\psi^{+ \hat{y}}_{+\frac{1}{2}+1}(x,\vec{k}_\perp)&\equiv & \frac{1}{\sqrt{2}}[\psi^{\uparrow}_{\frac{1}{2}+1}(x,\vec{k}_\perp)+ i \psi^{\downarrow}_{\frac{1}{2}+1}(x,\vec{k}_\perp)],\nonumber\\
\psi^{+ \hat{y}}_{+\frac{1}{2}-1}(x,\vec{k}_\perp)&\equiv & \frac{1}{\sqrt{2}}[\psi^{\uparrow}_{\frac{1}{2}-1}(x,\vec{k}_\perp)+ i \psi^{\downarrow}_{\frac{1}{2}-1}(x,\vec{k}_\perp)],\nonumber\\
\psi^{+ \hat{y}}_{-\frac{1}{2}+1}(x,\vec{k}_\perp)&\equiv & \frac{1}{\sqrt{2}}[\psi^{\uparrow}_{-\frac{1}{2}+1}(x,\vec{k}_\perp)+ i \psi^{\downarrow}_{-\frac{1}{2}+1}(x,\vec{k}_\perp)],\nonumber\\
\psi^{+ \hat{y}}_{-\frac{1}{2}-1}(x,\vec{k}_\perp)&\equiv & \frac{1}{\sqrt{2}}[\psi^{\uparrow}_{-\frac{1}{2}-1}(x,\vec{k}_\perp)+ i \psi^{\downarrow}_{-\frac{1}{2}-1}(x,\vec{k}_\perp)].\nonumber\\
\label{e12}
\eea
Using eqs. (\ref{spinup}) and (\ref{spindown}), we have
\bea
\psi^{+ \hat{y}}_{+\frac{1}{2}+1}(x,\vec{k}_\perp)&=&\frac{k^x-i k^y}{x(1-x)} \varphi, \nonumber\\
\psi^{+ \hat{y}}_{+\frac{1}{2}-1}(x,\vec{k}_\perp)&=& - \left(\frac{k^x+ i k^y}{1-x}  + i \left(M-\frac{m}{x}\right)\right)\varphi, \nonumber\\
\psi^{+ \hat{y}}_{-\frac{1}{2}+1}(x,\vec{k}_\perp)&=& - \left(\left(M-\frac{m}{x}\right) + i \frac{-k^x+ i k^y}{1-x}\right) \varphi, \nonumber\\
\psi^{+ \hat{y}}_{-\frac{1}{2}-1}(x,\vec{k}_\perp)&=& - i \frac{k^x + i k^y}{x (1-x)} \varphi .
\label{e13}
\eea
The unintegrated momentum space distribution, which is even in $k_\perp$, can be obtained by squaring the above equations
\bea
q_{\hat{y}}(x,\vec{k}_\perp)&=&\frac{1}{4 \pi}\left[|\psi^{+ \hat{y}}_{+\frac{1}{2}+1}(x,\vec{k}_\perp)|^2+|\psi^{+ \hat{y}}_{+\frac{1}{2}-1}(x,\vec{k}_\perp)|^2+|\psi^{+ \hat{y}}_{-\frac{1}{2}+1}(x,\vec{k}_\perp)|^2+|\psi^{+ \hat{y}}_{-\frac{1}{2}-1}(x,\vec{k}_\perp)|^2\right] \nonumber\\
&=& \frac{1}{2 \pi}\left[\frac{k^2_\perp (1+x^2)}{x^2 (1-x)^2}+\left(M-\frac{m}{x}\right)^2\right] \varphi^2 ,
\label{e14}
\eea
where $\varphi$ is supposed to be real.
We can prove explicitly that there is an asymmetry in the $\hat{x}$ direction in the state corresponding to eq. (\ref{e13}). For this purpose, we perform a Fourier transformation to the transverse position space coordinate say $\vec{f}_\perp$. We now have
\bea
\psi^{+ \hat{y}}_{+\frac{1}{2}+1}(x,\vec{f}_\perp)&\equiv& \int \frac{d^2 k_\perp}{(2 \pi)^2} e^{i \vec{f}_\perp \cdot \vec{k}_\perp} \psi^{+ \hat{y}}_{+\frac{1}{2}+1}(x,\vec{k}_\perp) \nonumber\\
&=& \frac{1}{x(1-x)}\left(- i \frac{\partial}{\partial f^x}-\frac{\partial}{\partial f^y}\right) \varphi(f_\perp) ,
\eea
\bea
\psi^{+ \hat{y}}_{+\frac{1}{2}-1}(x,\vec{f}_\perp)&\equiv& \int \frac{d^2 k_\perp}{(2 \pi)^2} e^{i \vec{f}_\perp \cdot \vec{k}_\perp} \psi^{+ \hat{y}}_{+\frac{1}{2}-1}(x,\vec{k}_\perp) \nonumber\\
&=& -\left[\frac{1}{(1-x)}\left(- i \frac{\partial}{\partial f^x}+\frac{\partial}{\partial f^y}\right)+ i \left(M-\frac{m}{x}\right)\right]\varphi(f_\perp) ,
\eea
\bea
\psi^{+ \hat{y}}_{-\frac{1}{2}+1}(x,\vec{f}_\perp)&\equiv& \int \frac{d^2 k_\perp}{(2 \pi)^2} e^{i \vec{f}_\perp \cdot \vec{k}_\perp} \psi^{+ \hat{y}}_{-\frac{1}{2}+1}(x,\vec{k}_\perp) \nonumber\\
&=& -\left[\left(M-\frac{m}{x}\right)+\frac{i}{1-x}\left(i \frac{\partial}{\partial f^x}+\frac{\partial}{\partial f^y}\right)\right]\varphi(f_\perp) ,
\eea
\bea
\psi^{+ \hat{y}}_{-\frac{1}{2}-1}(x,\vec{f}_\perp)&\equiv& \int \frac{d^2 k_\perp}{(2 \pi)^2} e^{i \vec{f}_\perp \cdot \vec{k}_\perp} \psi^{+ \hat{y}}_{-\frac{1}{2}-1}(x,\vec{k}_\perp) \nonumber\\
&=& \frac{- i}{x(1-x)} \left(-i \frac{\partial}{\partial f^x}+\frac{\partial}{\partial f^y}\right)\varphi(f_\perp) ,
\eea
where
\bea
\varphi(\vec{f}_\perp)&\equiv & \int \frac{d^2 k_\perp}{(2 \pi)^2} e^{i \vec{f}_\perp \cdot \vec{k}_\perp} \varphi(\vec{k}_\perp) \nonumber\\
&=& - e x \sqrt{1-x}  \int \frac{d^2 k_\perp}{(2 \pi)^2} e^{i \vec{f}_\perp \cdot \vec{k}_\perp} \frac{1}{k^2_\perp + C} \nonumber\\
&=& -\frac{e}{2 \pi} x \sqrt{1-x} K_0(|f_\perp \sqrt{C}|),
\eea
and
\be
C=m^2(1-x)- M^2 x (1-x)+ \lambda^2 x,
\ee
\begin{figure}
\minipage{0.42\textwidth}
   \includegraphics[width=6cm]{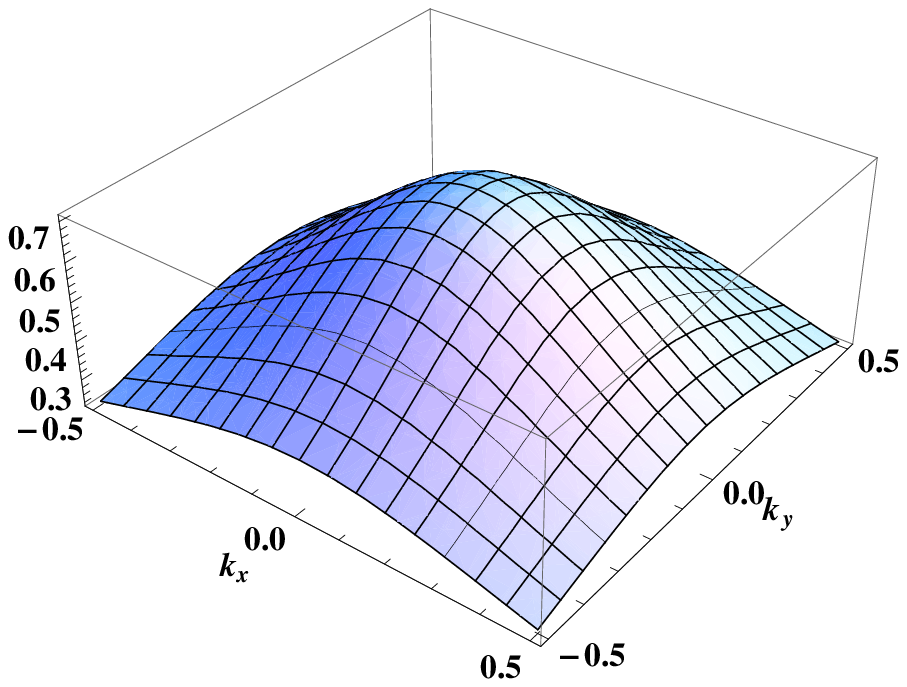}
  \endminipage\hfill
  \minipage{0.42\textwidth}
  \includegraphics[width=6cm]{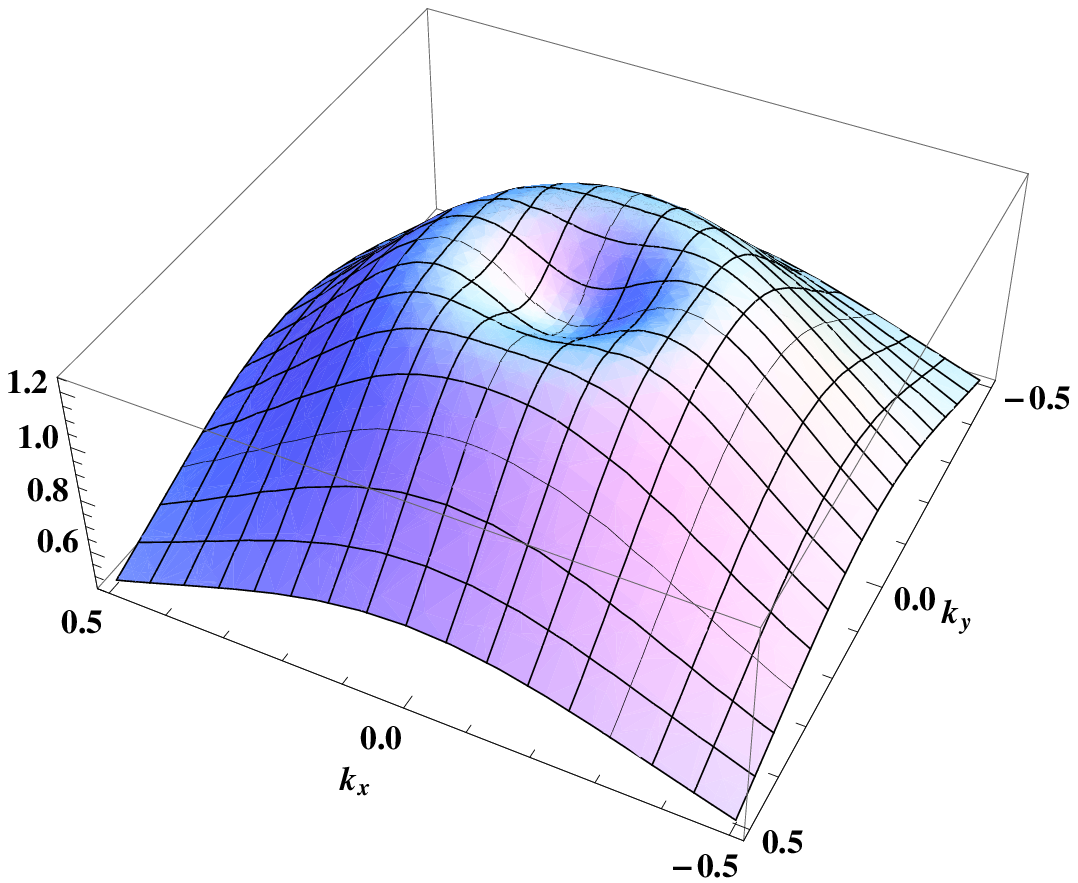}
  \endminipage\hfill
  \minipage{0.42\textwidth}
  \includegraphics[width=6cm]{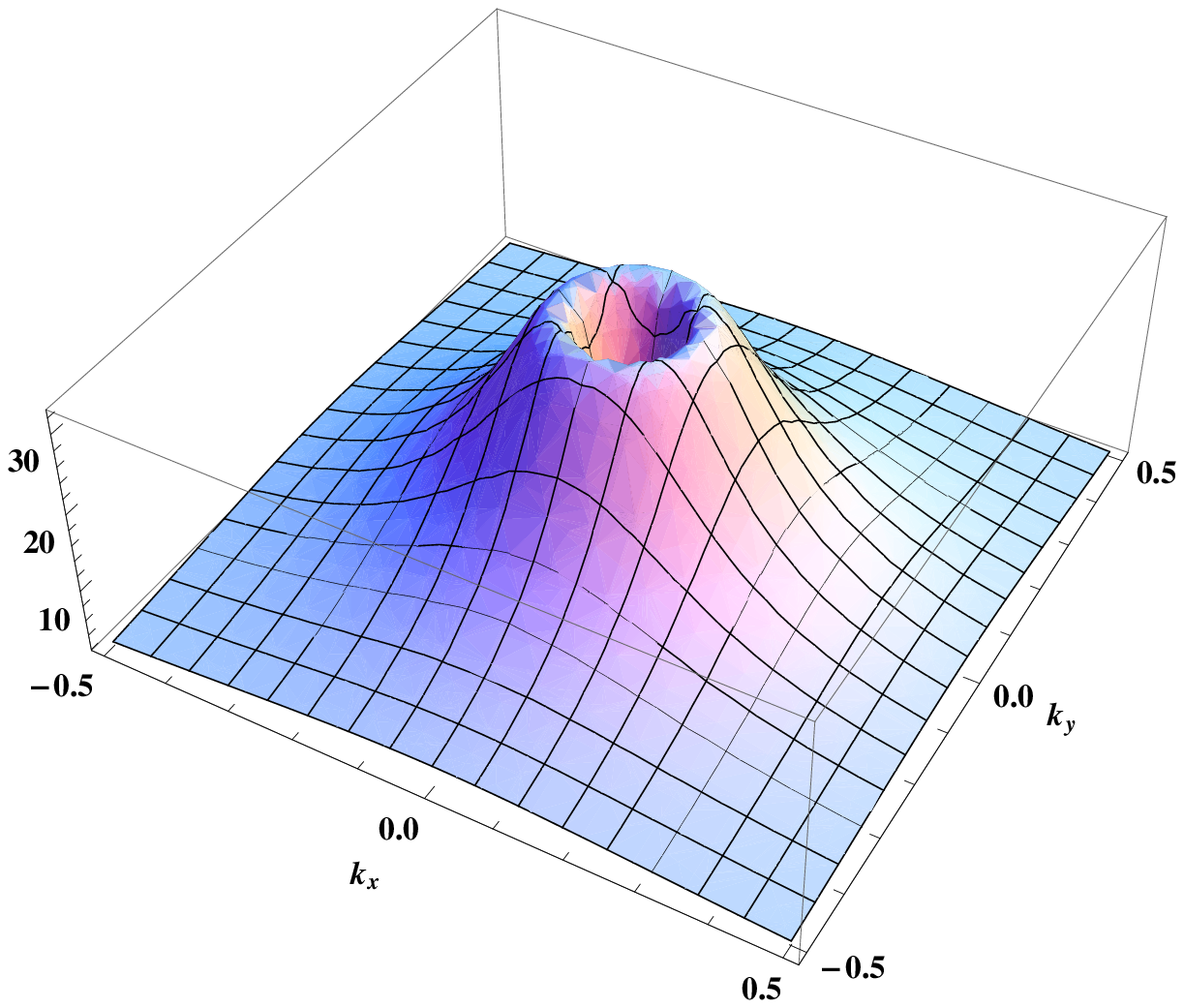}
  \endminipage\hfill
  \caption{Plot of $q_{\hat{y}}(x,\vec{k}_\perp)$ vs $k_\perp$ for three different values of $x=(0.1,0.4,0.8)$ .}
  \label{3d-distribution}
\end{figure}
Using the relations
\begin{eqnarray*}
|\psi^{+ \hat{y}}_{+\frac{1}{2}+1}(x,\vec{f}_\perp)|^2 &=& |\psi^{+ \hat{y}}_{-\frac{1}{2}-1}(x,\vec{f}_\perp)|^2, \nonumber\\
|\psi^{+ \hat{y}}_{+\frac{1}{2}-1}(x,\vec{f}_\perp)|^2 &=& |\psi^{+ \hat{y}}_{-\frac{1}{2}+1}(x,\vec{f}_\perp)|^2,
\label{e15}
\end{eqnarray*}
the unpolarized quark distribution in transverse coordinate space $\vec{f}_\perp$ can be expressed as
\bea
q_{\hat{y}}(x,\vec{f}_\perp)&=&\frac{1}{4 \pi}\left[|\psi^{+ \hat{y}}_{+\frac{1}{2}+1}(x,f_\perp)|^2+|\psi^{+ \hat{y}}_{+\frac{1}{2}-1}(x,f_\perp)|^2\right] \nonumber\\
&=&\frac{1}{2}\left[\frac{(1+x^2)}{x^2 (1-x)^2}\left(\frac{\partial}{\partial f^x} \varphi \right)^2+ \frac{1+x^2}{x^2 (1-x)^2}\left(\frac{\partial}{\partial f^y} \varphi\right)^2+ \left(M-\frac{m}{x}\right)^2 \varphi^2\right]\nonumber\\
&&-\left(M-\frac{m}{x}\right)\varphi \frac{1}{1-x}\left(\frac{\partial}{\partial f^x} \varphi \right).
\label{quark_distribution}
\eea
From this equation we observe that the last term is odd under $f^x \rightarrow - f^x$ and it defines the deformation in transverse coordinate space. This deformation is in agreement with the deformation predicted in eq. \ref{e9} in the impact parameter space. One can explicitly show the relation between the deformations in both the spaces using the convolution integrals. Since the convolution integrals are diagonalized by the FT, one can shift from the impact parameter space to the transverse position space representation as already shown. The transverse momentum $k_\perp$ in the two-particle Fock component is the Fourier conjugate of the distance $\vec{f}_\perp = \vec{r}_{\perp 1} - \vec{r}_{\perp 2}$ between the active quark and spectator system. The two variables $\vec{b}_\perp$ (transverse distance between the active quark and the center of mass momentum) and $\vec{f}_\perp$ can be related to each other by the relation $\vec{b}_\perp = (1-x) \vec{f}_\perp$ \cite{kumar2} and one can write
\bea
\frac{-i \frac{\partial}{\partial b^x}- \frac{\partial}{\partial b^y}}{2 M} \mathcal{E}(x,\vec{b}_\perp) &=& \frac{1}{4 \pi} \Big[\psi_{+\frac{1}{2}+1}^{\uparrow *}(x,\vec{f}_\perp) \psi_{+\frac{1}{2}+1}^{\downarrow}(x,\vec{f}_\perp)+\psi_{+\frac{1}{2}-1}^{\uparrow *}(x,\vec{f}_\perp) \psi_{+\frac{1}{2}-1}^{\downarrow}(x,\vec{f}_\perp) \Big] \frac{1}{(1-x)^2},\nonumber\\
\frac{\Delta^1 - i \Delta^2}{2 M} \mathcal{E}(x,b_\perp)&=& \frac{1}{4 \pi} \Bigg[ \frac{2 \varphi (k^1 -i k^2)}{1-x} \Bigg(M- \frac{m}{x}\Bigg) \varphi \Bigg] \frac{1}{(1-x)^2},\nonumber\\
\frac{1}{2M} \frac{\partial}{\partial b^x} \mathcal{E}(x, b_\perp) &=& \frac{2}{4 \pi (1-x)^2}\frac{1}{(1-x)} \Bigg(M-\frac{m}{x}\Bigg) \varphi \frac{\partial}{\partial f^x}\varphi.
\eea

In fig. \ref{3d-distribution} we present the unintegrated momentum space distribution obtained from the light front wavefunctions for different values of $x$, to check the sign of distortion for a polarized nucleon in impact parameter space. It is clear from the plots that at $x=0.1$ there is maxima at origin but as the value of $x$ is increased some distortion is observed. It can be easily seen from the plot that as the value at $x$ is further increased the distortion also increases towards negative direction. These plots helps us to determine the distortion sign directly from the light front wavefunctions.
\section{Conclusions}
In this present work we have studied the GPDs in impact parameter space obtained from LFWFs. We consider the spin-$\frac{1}{2}$ system consist of spin-$\frac{1}{2}$ fermion and spin-1 vector boson. We have shown that if spin flip GPD is non-zero then parton distribution is distorted in the transverse plane when the target nucleon has transverse polarization. We have obtained the  sign of the distortion from the sign of anomalous magnetic moment and our results are in agreement with the expected results.

Since the LFWFs can also be used directly to find the sign of distortion in impact parameter space, we have performed a FT of LFWFs in position space coordinate $\vec{f}_\perp$ and then explicitly shown the relation between the deformation obtained from GPDs in the impact parameter space and the deformation calculated directly from the LFWFs in the position space coordinate using the convolution integrals. These relations will also provide insight into the phenomena of shifting from the impact parameter space to the transverse position space representation. We consider the nucleon polarized in $+\hat{y}$ direction and then obtain the unintegrated momentum space distribution which is even in $k_\perp$. The deformation obtained in the impact parameter space is in agreement with the deformation predicted in transverse position space.

We have  designed a specific weight function of our model LFWFs and integrated it over the mass parameter to relate the LFWFs in the two-particle Fock state of the electron in QED to a realistic model of nucleon physics. The simulated Dirac and Pauli form factors obtained from LFWFs fall off at large $q^2$.
In addition to this,  we have simulated the form factor of the nucleon in the  AdS/QCD holographic LFWFs model and studied the power-law behaviour of wavefunction at short distances.  The magnitude of the form factor falls-off at large value of $q^2$. The light cone composite model used in the present work matches the power-law fall-off of form factors in perturbative QCD.

\section{Acknowledgement}
Authors acknowledge helpful discussion with S.J. Brodsky. HD would like to thank Department of Science and Technology (Ref No. SB/S2/HEP-004/2013), Government of India for financial support.

\end{document}